\documentclass[12pt]{article}

\usepackage{amsfonts}
\usepackage{amssymb}
\usepackage{amsmath}
\usepackage{amsthm}
\usepackage{indentfirst}
\usepackage{epsfig}
\usepackage{graphicx}
\usepackage{mathrsfs}
\usepackage{natbib}
\usepackage{bm, url}
\usepackage{epic}
\usepackage{curves}

\newtheorem{theorem}{Theorem}[section]
\newtheorem{assw}{Assumption}
\newtheorem{proposition}[theorem]{Proposition}
\newtheorem{lemma}[theorem]{Lemma}
\newtheorem{corol}[theorem]{Corollary}

\newtheorem{exam}{Example}[section] 

\newtheorem{rema}{Remark}[section] 

\newtheorem{defin}{Definition}[section] 

\newtheorem{ass}[assw]{Assumption}

\newenvironment{proofw}{\noindent {\bf Proof.}}{\hfill $\Box$\\}

\makeatletter

\numberwithin{equation}{section} %
\numberwithin{figure}{section}

\begin{document}

\title{Congestion, equilibrium and learning: \\ The minority game\footnote{Financial support from the Netherlands Organization for Scientific Research (NWO) and the Wallander/Hedelius Foundation is gratefully acknowledged.}}

\author{Willemien Kets\thanks{Corresponding author. Dept. of Econometrics
and Operations Research, Tilburg University, The Netherlands.
Address: Tilburg University, P.O. Box 90153, 5000 LE Tilburg, The
Netherlands. E-mail: w.kets@uvt.nl. Tel: +31-13-4662478. Fax:
+31-13-4663280.} \and Mark Voorneveld\thanks{Dept. of Economics,
Stockholm School of Economics, Sweden and Dept. of Econometrics
and Operations Research, Tilburg University, The Netherlands.}}
\date{August 2007}

\maketitle



\begin{abstract}
The minority game is a simple congestion game in which the
players' main goal is to choose among two options the one that is
adopted by the smallest number of players. We characterize the set
of Nash equilibria and the limiting behavior of several well-known
learning processes in the minority game with an arbitrary odd
number of players. Interestingly, different learning processes
provide considerably different predictions.
\end{abstract}



\thispagestyle{empty}

\newpage

\section{Introduction}

Congestion games are ubiquitous in economics. In a congestion game
\citep{Rosenthal1973}, players use several facilities from a
common pool. The costs or benefits that a player derives from a
facility depends on the number of users of that facility. A
congestion game is therefore a natural game to model scarcity of
common resources. Examples of such systems include vehicular
traffic \citep{NagelRasmussenBarrett1997}, packet traffic in
networks \citep{HubermanLukose1997}, and ecologies of foraging
animals \citep{DeAngelisGross1992}. Similar coordination problems
are encountered in market entry games \citep{SeltenGuth1982}.

Congestion games are also interesting from a theoretical point of
view. In congestion games, players need to coordinate to
differentiate. This seems to be more difficult than coordinating
on the same action, as any commonality of expectations is broken
up. For instance, when commuters have to choose between two roads
$A$ and $B$ and all believe that the others will choose road $A$,
nobody will choose that road, invalidating beliefs. The sorting of
players predicted in the pure-strategy Nash equilibria of such
games violates the common belief that in symmetric games, all
rational players will evaluate the situation identically, and
hence, make the same choices in similar situations \citep[see][p.
73]{HarsanyiSelten1988}. Moreover, in congestion games, players
may obtain asymmetric payoffs in equilibrium which may complicate
attainment of equilibrium, as coordination cannot be achieved
through tacit coordination based on historical precedent
\citep[cf.][]{Meyer_et_al_1992}. Finally, congestion games often
have many equilibria, so that players also face the difficulty of
coordinating on the same equilibrium.

Therefore, it is an interesting question what type of behavior
game theory predicts in such games. In this paper, we characterize
the equilibria of the minority game, a simple congestion game
based on the El Farol bar problem of \citet{Arthur1994}, and we
study the limiting behavior of a number of well-known learning
processes for this game. In the minority game, an odd number of
players
--- to make minorities well-defined --- choose between two ceteris
paribus identical alternatives. Congestion is costly, so players
prefer the alternative chosen by the smallest number of players.
The minority game is thus closely related to the market entry
game, a game extensively studied in experimental economics (see
the survey of \citet{Ochs1999} and references therein; for a
recent contribution see \citet{DuffyHopkins2005}). While the
market entry game models situations in which players can choose
between a safe option (staying out of the market) and an
alternative whose payoffs are declining in the number of other
players choosing that option (entering), the minority game is a
suitable model for more symmetric situations in which the payoffs
of both actions depend on the number of other players choosing
that action. In such situations, players will need to outsmart
other players, so as to be one step ahead of their opponents. For
instance, the minority game may be a good model for financial
markets, where investors try to identify the underpriced shares,
and try to sell the shares they expect to fall in the future. The
minority game has been studied by a number of authors in
economics. \citet{RenaultScarlattiScarsini2005} studies repeated
play in the game. \citet{BottazziDevetag2007},
\citet{ChmuraPitz2006}, and \citet{HelbingSchoenhofStark2005}
study the game experimentally. The game has also been studied
extensively in the physics literature; see
\citet{ChalletMarsiliZhang2004} or \citet{Coolen2005} for an
overview.

Interestingly, we find that the predictions from different
learning processes are not equivocal. While the replicator dynamic
predicts that play converges to a Nash equilibrium with at most
one player who chooses a strictly mixed strategy, the set of
stationary points under the perturbed best-response dynamics
consists of the logit quantal response equilibria of the game
\citep{McKelveyPalfrey1995}.\footnote{For a definition of these
learning processes, see Section~\ref{sec:repl_dyn} and
\ref{sec:perturbed_br_dyn}, respectively.} For the case of three
players, we show that the set of Nash equilibria that are the
limit of a sequence of logit quantal response equilibria with
vanishing noise consists of Nash equilibria with at most one mixer
and the Nash equilibrium in which all players randomize equally
over their two actions. Finally, we study the best-reply learning
process with limited memory of \citet{Hurkens1995} and the related
model of \citet{KetsVoorneveld2005}. \citeauthor{Hurkens1995}
studies a learning model in which players choose an arbitrary
action that is a best reply to some belief over other players'
actions that is consistent with their recent past play. In the
learning model of \citeauthor{KetsVoorneveld2005}, players also
best-reply to beliefs over others' play supported by recent past
play, but in addition, players additionally display a so-called
recency bias: when there are multiple best replies to a given
belief, a player chooses the action that he most recently played.
We show that while the process of \citeauthor{Hurkens1995} offers
no sharp predictions for the minority game, the model of
\citeauthor{KetsVoorneveld2005} predicts that play converges to
one of the pure Nash equilibria of the game when players have a
memory length of at least two periods.

\medskip

\noindent The current paper is related to the literature on
learning in congestion games and more generally learning in
potential games \citep[e.g.][]{HofbauerHopkins2005,
HofbauerSandholm2002,Sandholm2001, Sandholm2007}. Papers that
study learning in games similar to the game considered here
include \citet{Blonski1999}, \citet{Franke2003} and
\citet{KojimaTakahashi2004}. Most of these papers focus on the
predictions of a single learning
model,\footnote{\citet{DuffyHopkins2005} and
\citet{KojimaTakahashi2004} are notable exceptions.} while we
compare predictions from different learning models. Moreover,
while most results are obtained for games with either a small
number or a continuum of players, we characterize the equilibria
of the game and the limiting behavior of different learning
processes for any (odd) number of players.

The outline of this paper is as follows. In
Section~\ref{sec:game}, we define the game and characterize its
Nash equilibria. In Section~\ref{sec:repl_dyn}, we characterize
the set of stationary states and the set of asymptotically stable
states under the replicator dynamic. In
Section~\ref{sec:perturbed_br_dyn}, we characterize the set of
stationary states under the perturbed best-response dynamics.
In Section~\ref{sec:L2BP}, we characterize the limiting behavior
in the minority game under the best-reply learning processes with
limited memory. Section~\ref{sec:concl} concludes.

\section{The minority game} \label{sec:game}

\subsection{Basic definitions} \label{subsec:sym and mon}

Following the notation of \citet{TercieuxVoorneveld2005}, we
denote the set of players by $N = \{1, \dots, 2k + 1\}$, with $k
\in \mathbb{N}$. Each player $i \in N$ has a set of pure
strategies $A_i = \{-1,+1\}$: agents have to choose between two
options.  The set of mixed strategies of player $i$ is denoted by
$\Delta(A_i)$. We denote a mixed strategy profile by $\alpha \in
\times_{i \in N} \Delta(A_i)$, and we use the standard notation
$\alpha_{-i} \in \times_{j \in N \setminus \{i\}} \Delta(A_j)$ to
denote a strategy profile of players other than $i \in N$. With
each action $a \in \{-1,+1\}$, a function
\begin{equation*}
f_a : \{1, \dots, 2k+1\} \to \mathbb{R}
\end{equation*}
can be associated which indicates for each $n \in \{1, \dots,
2k+1\}$ the payoff $f_a(n)$ to a player choosing $a$ when the
total number of players choosing $a$ equals $n$. The von
Neumann-Morgenstern utility function of a player is then given by
\begin{equation}\label{eq:utility}
u_i(a) = f_{a_i}\left(|\{j \in N : a_j = a_i \}|\right),
\end{equation}
where $a \in \times_{j \in N} A_j$. Payoffs are extended to mixed
strategies in the usual way.

The function $f_a(\cdot), a \in \{-1,+1\}$ can have several forms.
We make the common assumptions
\citep[e.g.][]{ChalletMarsiliZhang2004} that congestion is costly:
\begin{quote}
\textbf{[Mon]} \qquad $f_{-1}$ and $f_{+1}$ are strictly
decreasing functions,
\end{quote}
and that the congestion effect is the same across alternatives:
\begin{quote}
\textbf{[Sym]} \qquad $f_{-1} = f_{+1}$.
\end{quote}
We refer to a player who uses a mixed strategy that puts positive
probability on both pure strategies a \emph{mixer}. A player that
puts full probability mass on the alternative $-1$ is called a
\emph{$(-1)$-player}; similarly, a player that puts full
probability mass on the alternative $+1$ is called a
\emph{$(+1)$-player}.

\subsection{Nash equilibria} \label{sec:equilibrium}

Throughout this section, let $k \in \mathbb{N}$ and consider a
minority game with $2k+1$ players. We characterize its set of Nash
equilibria. The pure Nash equilibria are easy to characterize:

\begin{proposition} \label{prop:pure_NE}{\rm \textbf{[\citet{TercieuxVoorneveld2005}]}} A pure
strategy profile is a Nash equilibrium if and only if one of the
alternatives $-1$ or $+1$ is chosen by exactly $k$ of the $2k+1$
players.
\end{proposition}

It remains to characterize the game's Nash equilibria with at
least one mixer.
\begin{lemma} \label{lemma:identical}
Let $\alpha \in \times_{i \in N} \Delta(A_i)$ be a Nash
equilibrium with a nonempty set of mixers. All mixers use the same
strategy: for all $i, j \in N$, if $\alpha_i, \alpha_j \notin
\{(1,0), (0,1)\}$, then $\alpha_i = \alpha_j$.
\end{lemma}
\begin{proofw}
By [Sym], the $2 \times 2$ subgame played by two mixers $i$ (row
player) and $j$ (column player) given the strategy profile of the
remaining players is of the form
\begin{center}
\begin{tabular}{ccc}
& $-1$ & $+1$ \\ \cline{2-3} $-1$ &
\multicolumn{1}{|c}{$x, x$} & \multicolumn{1}{|c|}{$y, z$} \\
\cline{2-3} $+1$ &
\multicolumn{1}{|c}{$z, y$} & \multicolumn{1}{|c|}{$w, w$} \\
\cline{2-3}
\end{tabular}
\end{center}
where, for instance, $y$ is the payoff to the player choosing $-1$
if the other player chooses $+1$ and the remaining players stick
to the mixed strategy profile $(\alpha_k)_{k \in N\setminus
\{i,j\}}$. By [Mon], a player is better off if the other chooses
differently, i.e., $x < y$ and $z > w$. Let $p, q \in (0,1)$
denote the equilibrium probability with which player $i$ and $j$,
respectively, choose $-1$. In equilibrium, each player must be
indifferent between playing $+1$ and playing $-1$:
\begin{eqnarray*}
p x + (1-p) y &=& p z + (1-p) w,\\
q x + (1-q) y &=& q z + (1-q) w.
\end{eqnarray*}
Subtracting the latter expression from the former yields
\begin{equation*}
(p-q)(x - y) = (p-q)(z - w).
\end{equation*}
As $x < y$ and $z > w$, this can only hold if $p = q$. Since
mixers $i$ and $j$ were chosen arbitrarily from the set of mixers,
this implies that all mixers use the same strategy.
\end{proofw}

Since all mixers use the same strategy and player labels are
irrelevant by [Sym] (if $\alpha$ is a Nash equilibrium, so is
every permutation of $\alpha$), a non-pure Nash equilibrium can be
summarized by its \emph{type} $(\ell, r, \lambda)$, where $\ell, r
\in \{0, 1, \ldots, 2k+1\}$ denote the number of players choosing
pure strategy $-1$ or $+1$, respectively, and $\lambda \in (0,1)$
the probability with which the remaining $m(\ell, r, \lambda) :=
(2k + 1) - (\ell + r) > 0$ mixers choose $-1$. Moreover, let
$v_{-1}(\ell, r, \lambda)$ denote the expected payoff to a player
choosing $-1$; $v_{+1}(\ell, r, \lambda)$ is defined similarly.
For convenience, write $m := m(\ell, r, \lambda)$. Letting one of
the mixers in $(\ell, r, \lambda)$ deviate to a pure strategy,
this implies in particular that
\begin{eqnarray}
v_{-1}(\ell + 1, r, \lambda) & = & \sum_{s = 0}^{m-1}
\binom{m-1}{s} \lambda^s (1 -
\lambda)^{m-1-s} f_{-1}(\ell + 1 + s), \label{eq:vminus} \\
v_{+1}(\ell, r+1, \lambda) & = & \sum_{s = 0}^{m-1} \binom{m-1}{s}
\lambda^s (1 - \lambda)^{m-1-s} f_{+1}((r+1) + (m-1-s)) \nonumber \\
& = & \sum_{s = 0}^{m-1} \binom{m-1}{s} \lambda^s (1 -
\lambda)^{m-1-s} f_{+1}(r + m - s). \label{eq:vplus}
\end{eqnarray}
For instance, a profile of type $(\ell + 1, r, \lambda)$ is
obtained from type $(\ell, r, \lambda)$ if a mixer switches to
pure strategy $-1$. In that case, there are $m -1$ mixers left. To
obtain expected payoffs, notice that the probability that $s \in
\{0, \ldots, m-1\}$ of these mixers choose $-1$ is $\binom{m-1}{s}
\lambda^s (1 - \lambda)^{m-1-s}$. Using this notation, the Nash
equilibria with at least one mixer are characterized as follows.

\begin{proposition} \label{prop:charact mixers} \quad

\begin{description}
\item[(a) (Characterization of equilibrium)] Let $\ell, r \in \{0,
1, \ldots, 2k+1\}$ be such that $\ell + r < 2k + 1$. Let $\lambda
\in (0,1)$. A strategy profile of type $(\ell, r, \lambda)$ is a
Nash equilibrium if and only if
\begin{equation} \label{eq:mixer indif}
v_{-1}(\ell + 1, r, \lambda) = v_{+1}(\ell, r+1, \lambda).
\end{equation}

\item[(b) (Equilibria with one mixer)] There exist equilibria with
exactly one mixer. These equilibria are of type $(k,k, \lambda)$
with arbitrary $\lambda \in (0,1)$, i.e., the mixer uses an
arbitrary mixed strategy, whereas the remaining $2k$ players are
spread evenly over the two pure strategies.

\item[(c) (Equilibria with more than one mixer)] Let $\ell, r \in
\{0, 1, \ldots, 2k+1\}$ be such that $\ell + r \leq 2k - 1$. There
is a Nash equilibrium of type $(\ell, r, \lambda)$ if and only if
$\max\{\ell, r\} < k$. The corresponding probability $\lambda \in
(0,1)$ solving \eqref{eq:mixer indif} is unique.
\end{description}
\end{proposition}
\begin{proofw}
\noindent \textbf{(a):} Condition \eqref{eq:mixer indif} says that
a mixer is indifferent between choosing $-1$, thereby raising
$\ell$ to $\ell + 1$ and obtaining payoff $v_{-1}(\ell + 1, r,
\lambda)$, or choosing $+1$, thereby raising $r$ to $r+1$ and
obtaining payoff $v_{+1}(\ell, r+1, \lambda)$. Hence,
\eqref{eq:mixer indif} is a necessary condition for Nash
equilibrium.

To establish sufficiency, it remains to show that also players
using a pure strategy
--- if there are such players, i.e., if $\ell + r \geq 1$ --- choose a best reply.
Suppose $\ell \geq 1$. The payoff to a $(-1)$-player is
$v_{-1}(\ell, r, \lambda)$, while a unilateral deviation to $+1$
yields $v_{+1}(\ell - 1, r + 1, \lambda)$. However:
\begin{eqnarray}
v_{-1}(\ell, r, \lambda) & \geq & v_{-1}(\ell+1, r, \lambda) \label{eq:smon1} \\
& = & v_{+1}(\ell, r+1, \lambda) \label{eq:use mixer indif} \\
& \geq & v_{+1}(\ell-1, r+1, \lambda). \label{eq:smon2}
\end{eqnarray}
Inequality \eqref{eq:smon1} uses [Mon]: conditioning on the
behavior of one of the $m := m(\ell, r, \lambda) > 0$ mixers,
write
\[
v_{-1}(\ell, r, \lambda) = \lambda v_{-1}(\ell + 1, r, \lambda) +
(1 - \lambda) v_{-1}(\ell, r+1, \lambda).
\]
Then
\begin{eqnarray*}
\lefteqn{v_{-1}(\ell,r, \lambda) - v_{-1}(\ell + 1, r, \lambda) =
(1- \lambda) \left[ v_{-1}(\ell, r+1, \lambda) - v_{-1}(\ell + 1,
r,
\lambda)\right]} \\
&& = (1- \lambda) \sum_{s=0}^{m-1}
\binom{m-1}{s}\lambda^s(1-\lambda)^{m-1-s} \left[ f_{-1}(\ell+s) -
f_{-1}(\ell+1+s) \right] \\
&& \geq 0
\end{eqnarray*}
by [Mon]. Inequality \eqref{eq:smon2} follows similarly and
\eqref{eq:use mixer indif} is simply condition \eqref{eq:mixer
indif}. So if $\ell \geq 1$, $(-1)$-players choose a best reply.
Similarly, if $r \geq 1$, $(+1)$-players choose a best reply.

\noindent \textbf{(b):} Let $\lambda \in (0,1)$. Substitution in
\eqref{eq:mixer indif} and [Sym] yield that strategy profiles of
type $(k,k, \lambda)$ are Nash equilibria:
\[
v_{-1}(k+1,k,\lambda) = f_{-1}(k+1) = f_{+1}(k+1) = v_{+1}(k, k+1,
\lambda).
\]
Conversely, consider a Nash equilibrium of type $(\ell, r,
\lambda)$ with exactly one mixer: $\ell + r = 2k$. We establish
that $\ell = r$. Suppose not. W.l.o.g., $\ell > r$. Since $\ell +
r = 2k$, this implies $\ell \geq k+1$ and $r \leq k-1$. The
expected payoff to a $(-1)$-player is
\[
\lambda f_{-1}(\ell + 1) + (1 - \lambda) f_{-1}(\ell),
\]
while deviating to $+1$ would yield
\[
\lambda f_{+1}(r + 1) + (1 - \lambda) f_{+1}(r + 2).
\]
Since $\ell + 1 > r + 1, \ell \geq r + 2$, and $\lambda \in
(0,1)$, it follows from [Sym] and [Mon] that a $(-1)$-player would
benefit from unilateral deviation, contradicting the assumption
that the profile of type $(\ell, r, \lambda)$ is a Nash
equilibrium. Conclude that $\ell = r$.

\noindent \textbf{(c):} Without loss of generality, $\ell \geq r$,
so $\max \{\ell, r\} = \ell$. Let $m = (2k+1) - (\ell + r) \geq 2$
be the number of mixers. By substitution, $\ell < k$ if and only
if $\ell + 1 < r + m$. To prove (c), it therefore remains to
establish three things.

Firstly, if $\ell + 1 < r + m$, there is a $\lambda \in (0,1)$
solving \eqref{eq:mixer indif}. To see this, use $\ell \geq r$ to
find that $\ell + m > r + 1$. By [Sym] and [Mon], it follows that
\[
\begin{array}{ccccccc}
v_{-1}(\ell + 1, r, 0) & = & f_{-1}(\ell + 1) & > & f_{+1}(r + m)
& = & v_{+1}(\ell, r
+1, 0), \\
v_{-1}(\ell + 1, r, 1) & = & f_{-1}(\ell + m) & < & f_{+1}(r + 1)
& = & v_{+1}(\ell, r +1, 1).
\end{array}
\]
By the Intermediate Value Theorem applied to $v_{-1}(\ell +1, r,
\cdot) - v_{+1}(\ell, r+1,\cdot)$, there is a $\lambda \in (0,1)$
solving \eqref{eq:mixer indif}: there is a Nash equilibrium of
type $(\ell, r, \lambda)$.

Secondly, this $\lambda \in (0,1)$ solving \eqref{eq:mixer indif}
is unique. By \eqref{eq:vminus}, $v_{-1}(\ell + 1, r, \cdot)$ is
the expectation of a strictly decreasing function of a binomial
stochastic variable. By stochastic dominance (see
Appendix~\ref{app:stoch_dom_binom}), this makes $v_{-1}(\ell + 1,
r, \cdot)$, the left-hand side of \eqref{eq:mixer indif}, strictly
decreasing in $\lambda$. Similarly, by \eqref{eq:vplus}, the
right-hand side of \eqref{eq:mixer indif} is strictly increasing
in $\lambda$. Conclude that the functions $v_{-1}(\ell + 1, r,
\cdot)$ and $v_{+1}(\ell, r+1, \cdot)$ intersect at most once. By
the previous step, as long as $\ell + 1 < r + m$, they intersect
at least once, establishing uniqueness.

Thirdly, if $\ell + 1 \geq r + m$, there is no $\lambda \in (0,1)$
solving \eqref{eq:mixer indif}. To see this, notice that the
inequality implies
\[
\ell + m > \cdots > \ell + 2 > \ell + 1 \geq r + m > r + m - 1 >
\cdots > r + 1,
\]
so by [Sym] and [Mon]:
\[
f_{-1}(\ell + m) < \cdots < f_{-1}(\ell + 2) < f_{-1}(\ell + 1)
\leq f_{+1}(r + m) < f_{+1}(r + m - 1) < \cdots < f_{+1}(r + 1).
\]
Substitution in \eqref{eq:vminus} and \eqref{eq:vplus} yields that
\[
v_{+1}(\ell, r+1, \lambda) > v_{-1}(\ell + 1, r, \lambda)
\]
for all $\lambda \in (0,1)$: there is no solution to
\eqref{eq:mixer indif}.
\end{proofw}

\medskip
\noindent Some consequences of this characterization of the game's
non-pure Nash equilibria:

\noindent \textbf{(i):} There are no Nash equilibria where the
number of mixers is two, since in that case, $\max \{\ell, r\}
\geq k$.

\noindent \textbf{(ii):} Substitution in \eqref{eq:mixer indif}
gives that a strategy profile in which the number of
$(-1)$-players is equal to the number of $(+1)$-players and the
remaining players mix with probability $1/2$, i.e., a profile of
type $(t, t, 1/2)$ with $t \in \{0, \ldots, k\}$, is a Nash
equilibrium.

\medskip
\noindent Having characterized the set of Nash equilibria, we now
establish that the set of Nash equilibria with at most one mixer
is connected.
\begin{proposition}
The set of Nash equilibria with at most one mixer is connected.
\end{proposition}
\begin{proofw}
In a Nash equilibrium with exactly one mixer, the completely mixed
strategy is arbitrary. Letting the probability go to zero or one,
this line piece of Nash equilibria in the strategy space has a
pure Nash equilibrium as its end point. Hence, to show
connectedness, it suffices to show that for each pair of pure Nash
equilibria, there is a chain of pure Nash equilibria differing in
exactly one coordinate connecting them.

So let $x$ and $y$ be distinct pure Nash equilibria. By
Proposition \ref{prop:pure_NE}, the majority action, i.e., the
action chosen by exactly $k+1$ players in a given Nash
equilibrium, is well-defined. We need to consider two cases.
Firstly, if this action is the same in $x$ and $y$, w.l.o.g. $-1$,
then $x \neq y$ implies that the $(k+1)$-player majorities in $x$
and $y$ must be distinct. Let $i$ be such a majority player,
choosing $-1$ in $x$, but $+1$ in $y$. Secondly, if the majority
action is different in $x$ and $y$, w.l.o.g. $-1$ in $x$ and $+1$
in $y$, then by definition of a majority, the $(k+1)$-player
majorities in $x$ and $y$ have a nonempty intersection. Again, let
$i$ be a majority player choosing $-1$ in $x$, but $+1$ in $y$.

By construction, as $i$ is a majority player, the path of Nash
equilibria in which $i$ increases the probability of playing the
action $+1$ from $0$ to $1$ connects $x$ to another pure Nash
equilibrium $x^*$ with $x_i \neq x^*_i = y_i$ and $x^*_j = y_j$
for all $j \neq i$, i.e., with a strictly smaller Hamming distance
to $y$ (recall that the Hamming distance between two
finite-dimensional vectors is the number of coordinates in which
they differ).

As the strategy vectors have only finitely many coordinates and we
can reduce the Hamming distance between pure Nash equilibria by
the procedure above, the result now follows by induction.
\end{proofw}

\section{The replicator dynamic} \label{sec:repl_dyn}

In this section, we study the replicator dynamic
\citep[e.g.][]{Weibull1995} for the minority game. There is a set
$N = \{1, \ldots, 2k+1\}$ of populations, where each population is
the unit interval $[0,1]$. The populations represent the $2k + 1$
player positions in the minority game. All agents in a population
are initially programmed to some pure strategy. Hence, each
population can be divided into two subpopulations (one of which
may contain no agents), one for each of the pure strategies in the
minority game. A \emph{population state} is a vector $\alpha =
(\alpha_1, \ldots, \alpha_{2k+1})$ in the polyhedron of
mixed-strategy profiles, where for each $i \in N$, $\alpha_i$ is a
point in the simplex $\Delta(A_i)$, representing the distribution
of agents in population $i$ across the different pure strategies.
The vector $\alpha_i \in \Delta(A_i)$ thus represents the state of
population $i$, with $\alpha_i(a_i)$ denoting the proportion of
agents programmed to play the pure strategy $a_i \in A_i$.

Time is continuous and indexed by $t$. Agents -- one from each
population -- are continuously drawn uniformly at random from
these populations to play the minority game. Suppose payoffs
represent the effect of playing the game on an agent's fitness,
measured as the number of offspring per time unit, and that each
offspring inherits its single parent's strategy. This gives rise
to the following dynamics for the population shares:
\begin{equation} \label{eq:repl u}
\forall i \in N, \forall a_i \in A_i: \quad \dot{\alpha_i}(a_i) =
\alpha_i(a_i) (u_i(a_i, \alpha_{-i}) - u_i(\alpha_i,
\alpha_{-i})).
\end{equation}
This system of differential equations defines the (continuous time
multipopulation) \emph{replicator dynamic}.
In words, the growth rate $\dot{\alpha_i}(a_i)/\alpha_i(a_i)$ of a
pure strategy $a_i \in A_i$ in population $i \in N$ is equal to
the difference in payoffs of the pure strategy and the current
average payoffs for the population. Hence, the population shares
of strategies that do better than average will grow, while the
shares of the other strategies will decline. It is easily seen
that the subpopulations associated with the pure best replies to
the current population state have the highest growth rates.

The system of differential equations~\eqref{eq:repl u} defines a
continuous \emph{solution mapping} $\xi:\mathbb{R} \times
\bigl(\times_{i \in N} \Delta(A_i)\bigr) \to \times_{i \in N}
\Delta(A_i) $ which assigns to each time $t \in \mathbb{R}$ and
each initial state $\alpha^0 \in \times_{i \in N} \Delta(A_i)$ the
population state $\xi(t,\alpha^0) \in \times_{i \in N}
\Delta(A_i)$. The (solution) \emph{trajectory} through a
population state $\alpha^0 \in \times_{i \in N} \Delta(A_i)$ is
the graph of the solution mapping $\xi(\cdot, \alpha^0)$.

A population state $\alpha \in \times_{i \in N} \Delta(A_i)$ is a
\emph{stationary state} of the replicator dynamics~\eqref{eq:repl
u} if and only if for each population $i \in N$ every pure
strategy $a_i \in A_i$ that is used by some agents in the
population gives the same payoffs. In that case,
$\dot{\alpha_i}(a_i) = 0$ for all $i \in N$ and all $a_i \in A_i$.
Let $S = \{\alpha \in \times_{j \in N} \Delta(A_j) \mid \forall i
\in N, \forall a_i \in A_i: \dot{\alpha_i}(a_i) = 0\}$ be the set
of stationary states. By definition, if $\alpha \in S$, then a
player $i \in N$ either uses a pure strategy or --- if he is a
mixer --- is indifferent between his two pure strategies:
$u_i(a_i, \alpha_{-i}) = u_i(\alpha_i, \alpha_{-i})$ for both $a_i
\in A_i$. Using the proof of Lemma \ref{lemma:identical}, all
mixers must use the same strategy. If there is more than one
mixer, the proof of Proposition \ref{prop:charact mixers}(c)
indicates that this mixed strategy solving \eqref{eq:mixer indif}
is uniquely determined by the number of players choosing pure
strategy $-1$ and pure strategy $+1$. Conclude that the set of
stationary states can be partitioned into three subsets:
\begin{itemize}
\item[$S_1$:] The connected set of Nash equilibria with at most
one mixer;
\end{itemize}
and a finite collection of isolated stationary states, namely
\begin{itemize}
\item[$S_2$:] Nash equilibria with more than one mixer;

\item[$S_3$:] nonequilibrium profiles of some type $(\ell, r,
\lambda)$, where
\[
\left\{
\begin{array}{l}
\ell, r \in \{0, \ldots, 2k+1\}, \\
\ell + r \leq 2k+1, \\
\mbox{if } \ell + r < 2k+1, \mbox{ then $\lambda \in (0,1)$
uniquely determined by \eqref{eq:mixer indif}.}
\end{array}
\right.
\]
\end{itemize}

It remains to study the stability properties of these stationary
states. We consider Lyapunov stability and asymptotic stability.
Roughly speaking, a population state is Lyapunov stable if no
small change in the population shares can lead the replicator
dynamics away from the population state, while a population state
is asymptotically stable if it is Lyapunov stable and any
sufficiently small change in the population shares results in a
movement back to the original population state. Formally, a
population state $\alpha \in \times_{i \in N}\Delta(A_i)$ is
\emph{Lyapunov stable} if every neighborhood $B$ of $\alpha$
contains a neighborhood $B^0$ of $\alpha$ such that
$\xi(t,\alpha^0) \in B$ for every $x^0 \in B \cap \times_{i \in
N}\Delta(A_i)$ and $t \geq 0$. It is \emph{asymptotically stable}
if it is Lyapunov stable, and, in addition, there exists a
neighborhood $B^*$ such that
$$\lim_{t \to \infty} \xi(t, \alpha^0) = \alpha$$ for each initial
state $\alpha^0 \in B^* \cap \times_{i \in N}\Delta(A_i)$.

The analysis relies heavily on the existence of a Lyapunov
function for the replicator dynamic in the minority game.
\citet{TercieuxVoorneveld2005}, using Thm. 3.1 in
\citet{MondererShapley1996}, show that a minority game is a
(finite exact) potential game: there exists a real-valued
(so-called potential) function $U$ on the pure strategy space such
that for each $i \in N$, each $a_{-i} \in \times_{j \in N\setminus
\{i\}} A_j$, and all $a_i, b_i \in A_i$:
\begin{equation} \label{eq:def potential}
u_i(a_i, a_{-i}) - u_i(b_i, a_{-i}) = U(a_i, a_{-i}) - U(b_i,
a_{-i}).
\end{equation}
Taking expectations, \eqref{eq:def potential} can be extended to
mixed strategies, so the payoff difference in \eqref{eq:repl u}
equals the corresponding change in the potential. Hence, the
replicator dynamic can be rewritten as:
\begin{equation} \label{eq:repl P}
\forall i \in N, \forall a_i \in A_i: \quad \dot{\alpha_i}(a_i) =
\alpha_i(a_i) (U(a_i, \alpha_{-i}) - U(\alpha_i, \alpha_{-i})).
\end{equation}
This makes the potential $U$ a Lyapunov function of the replicator
dynamic. More precisely:

\begin{proposition} \label{prop:lyapunov}
The potential function $U$ of the minority game is a strict
Lyapunov function for the replicator dynamic: for each solution
trajectory $(\alpha(t))_{t \in [0, \infty)}$, $dU(\alpha(t))/dt
\geq 0$ with equality exactly in the stationary states.
\end{proposition}
\begin{proofw}
Suppressing time indices for ease of notation, direct calculation
gives
\begin{eqnarray*}
\dot{U}(\alpha) & = & \sum_{i \in N} \sum_{a_i \in A_i} U(a_i,
\alpha_{-i})
\dot{\alpha_i}(a_i) \\
& = & \sum_{i \in N} \sum_{a_i \in A_i} \alpha_i(a_i) (U(a_i,
\alpha_{-i}) -
U(\alpha_i, \alpha_{-i})) U(a_i, \alpha_{-i}) \\
& = & \sum_{i \in N} \sum_{a_i \in A_i} \left(\alpha_i(a_i) U(a_i,
\alpha_{-i})^2 -
U(\alpha_i, \alpha_{-i})^2 \right)\\
& = & \sum_{i \in N} \left(\mathbb{E}_{\alpha_i} \left[ U(a_i,
\alpha_{-i})^2 \right] - \left(\mathbb{E}_{\alpha_i} \left[ U(a_i,
\alpha_{-i}) \right] \right)^2
\right) \\
& = & \sum_{i \in N} \mbox{Var}_{\alpha_i} U(a_i, \alpha_{-i}) \\
& \geq & 0,
\end{eqnarray*}
with equality if and only if all variances are zero, i.e., if and
only if $\alpha$ is a stationary point of the replicator dynamics.
\end{proofw}

\begin{proposition} \label{prop:repl_dyn_asympt_stab}
The collection of Nash equilibria with at most one mixer in $S_1$
is asymptotically stable under the replicator dynamic. Stationary
states in $S_2$ and $S_3$ are not Lyapunov stable.
\end{proposition}
\begin{proofw}
To see that the collection of Nash equilibria in $S_1$ is
asymptotically stable, notice that $S_1$ is the set of global
maxima of $U$: The potential $U$ in \eqref{eq:def potential} was
extended to mixed strategies by taking expectations, so $U$
achieves a global maximum in a pure strategy profile which, again
by \eqref{eq:def potential}, is a pure Nash equilibrium. By
symmetry, all pure Nash equilibria are global maxima of $U$ and so
are equilibria with exactly one mixer. Other strategy profiles are
not global maxima of $U$: they are not Nash equilibria or, if they
are, they involve more than one mixer, in which case they put
positive probability also on pure strategy profiles that are not
Nash equilibria and consequently not global maxima of $U$. This
connected set of global maxima of the Lyapunov function $U$ is
asymptotically stable \citep[][Thm. 6.4]{Weibull1995}.

We show that elements of $S_2$ are not Lyapunov stable; the case
for points in $S_3$ is similar. Let $\alpha^* \in S_2$, i.e,
$\alpha^*$ is a Nash equilibrium with more than one mixer. Suppose
it is Lyapunov stable. Since it is an isolated point of the
collection of stationary states, there is a neighborhood $B$ of
$\alpha^*$ whose closure contains only the stationary state
$\alpha^*$: $\mbox{cl} (B) \cap S_2 = \{\alpha^*\}$. By Lyapunov
stability, as long as the initial state $\alpha(0)$ lies in a
sufficiently small neighborhood $B'$ of $\alpha^*$, the entire
solution trajectory $(\alpha(t))_{t \in [0, \infty)}$ remains in
$B$.

Let $i \in N$ be one of the mixers in the Nash equilibrium
$\alpha^*$. Since $i$ is indifferent between his two pure
strategies and the potential $U$ measures payoff differences, it
follows that
\[
U(\alpha^*) = U(-1, \alpha^*_{-i}) = U(+1, \alpha^*_{-i}).
\]
Consequently, $U(\gamma_i, \alpha^*_{-i}) = U(\alpha^*)$ for all
mixed strategies $\gamma_i$ of player $i$. For $\gamma_i \neq
\alpha^*_i$ sufficiently close to $\alpha^*_i$, it follows that
$(\gamma_i, \alpha^*_{-i}) \in B'$. Hence, the entire solution
trajectory $(\gamma(t))_{t \in [0, \infty)}$ with $\gamma(0) :=
(\gamma_i, \alpha^*_{-i})$ remains in $B$. Since its starting
point is not stationary, Proposition \ref{prop:lyapunov} implies
that the Lyapunov function $U$ strictly increases along the
trajectory, until it may reach a stationary state. Let $\gamma^*
\in \times_{j \in N} \Delta(A_j)$ be a limit point of the
trajectory $(\gamma(t))_{t \in [0, \infty)}$: there is a strictly
increasing sequence of time points $t_m \rightarrow \infty$ such
that $\lim_{m \rightarrow \infty} \gamma(t_m) \rightarrow
\gamma^*$. Such a limit point exists and has to be a stationary
point \citep[Lemma A.1 of][p. 104]{Sandholm2001}. Since $\mbox{cl}
(B) \cap S_2 = \{\alpha^*\}$ and the trajectory lies in $B$, it
follows that $\gamma^* = \alpha^*$. But then $\lim_{m \rightarrow
\infty} U(\gamma(t_m)) = U(\alpha^*) = U(\gamma(0))$,
contradicting that the Lyapunov function is increasing along the
trajectory. Conclude that $\alpha^*$ cannot be Lyapunov stable.
For $\alpha^* \in S_3$, proceed similarly. As it is not a NE, some
$i$ can profitably deviate slightly (to remain inside $B'$), so
the remaining trajectory must increase the potential, but still
have $\alpha^*$ as its limit point.
\end{proofw}


\section{Perturbed best-response dynamics and quantal response equilibria}
\label{sec:perturbed_br_dyn}

\subsection{Perturbed best-response dynamics}

Under stochastic fictitious play
\citep[e.g.][]{HofbauerHopkins2005,HofbauerSandholm2002,
Hopkins2002}, players repeatedly play a normal form game (in
discrete time). They choose best replies to their beliefs on other
players' actions on the basis of a perturbed payoff function, with
beliefs determined by the time average of past play. More
specifically, the state variable at time $t \in \mathbb{N}$ is a
vector $Z^t \in \times_{i \in N} \Delta(A_i)$, where the $i$th
component $Z^t_i$ denotes the time average of player $i$'s past
play up to time $t$. Players' initial choices are arbitrary pure
strategies; in later periods players best-respond to their beliefs
$Z^t$, after their payoffs have been subjected to random shocks.
That is, for each $i \in N$, let $(\varepsilon_i^a)_{a \in A_i}$
be a vector of payoff disturbances. The vector of payoff
disturbances is independent and identically distributed across
players and over time.  Let $\alpha_{-i} \in \times_{j \in N
\setminus \{i\}} \, \Delta(A_j)$ be a belief. The probability that
player $i$ chooses action $a_i \in A_i$ is equal to the
probability that
$$u_i(a_i, \alpha_{-i}) + \varepsilon_i^{a_i} \geq u_i(b_i, \alpha_{-i}) +
\varepsilon_i^{b_i}$$ for all $b_i \in A_i$. Then, the perturbed
best-response dynamics associated with Gumbel-distributed
perturbations with parameter $\beta > 0$ is:
\begin{equation} \label{eq:PV}
\forall i \in N, \forall a_i \in A_i: \dot{\alpha}_i(a_i) =
\frac{\exp\left[\beta u_i(a_i, \alpha_{-i})\right]}{\sum_{b_i \in
A_i} \exp\left[\beta u_i(b_i, \alpha_{-i})\right]} -
\alpha_i(a_i).
\end{equation}
Gumbel-distributed payoff perturbations correspond to control
costs of the relative entropy form. By Proposition 4.1 of
\citet{HofbauerSandholm2002}, the process in \eqref{eq:PV} has a
strict Lyapunov function that can be expressed in terms of the
potential function and the control cost functions. For each $i \in
N$, let $\alpha_i$ denote the probability with which player $i$
chooses the action $a_i = -1$. Then, the Lyapunov function for the
process in \eqref{eq:PV} is defined by:
\begin{equation}\label{eq:lyapunov_pv}
\alpha \in \times_{i \in N} \Delta(A_i): \quad V(\alpha) =
U(\alpha) - \frac{1}{\beta} \sum_{i \in N} \left[ \alpha_i
\log(\alpha_i) + (1-\alpha_i) \log(1-\alpha_i)\right],
\end{equation}
where $U$ is the potential function.
%
Since control cost functions of the relative entropy form satisfy
the smoothness conditions of Proposition 4.2 of
\citet{HofbauerSandholm2002}, it follows that:
\begin{proposition}
The collection of stationary states and recurrent points of the
process in \eqref{eq:PV} coincide.
\end{proposition}
Theorem 6.1(iii) of \citet{HofbauerSandholm2002} now implies that
the perturbed best-response dynamic converges to these stationary
states. Notice that the set of stationary states coincides with
the set of logit quantal response equilibria of the minority game
\citep{McKelveyPalfrey1995}. When the perturbation terms go to
zero, we obtain Nash equilibria. As the set of Nash equilibria is
not finite, we cannot apply Corollary 6.6 of \citet{Benaim1999} to
characterize the subset of Nash equilibria to which the stochastic
process~\eqref{eq:PV} converges. The set of Nash equilibria that
are the limit points of a sequence of logit quantal response
equilibria is generally hard to characterize. In the next section,
we characterize this set for the three-player minority game.

\subsection{Stationary points for the three-player minority game}

Consider the three-player minority game with $f_{-1} = f_{+1} = f$
strictly decreasing in the number of users. As it involves a
simple rescaling of functions satisfying [Mon] and [Sym], we may
without loss of generality set $f(2) = 0$ and $f(1) - f(3)= 1$. A
potential of the game is then given in Figure~\ref{fig:potential}.
\begin{figure}[h]
\begin{center}
\begin{tabular}{ccc}
& $-1$ & $+1$ \\ \cline{2-3} $-1$ & \multicolumn{1}{|c}{$-1$} &
\multicolumn{1}{|c|}{$0$} \\ \cline{2-3} $+1$ &
\multicolumn{1}{|c}{$0$} & \multicolumn{1}{|c|}{$0$} \\
\cline{2-3}
\end{tabular} \quad
\begin{tabular}{ccc}
& $-1$ & $+1$ \\ \cline{2-3} $-1$ & \multicolumn{1}{|c}{$0$} &
\multicolumn{1}{|c|}{$0$} \\ \cline{2-3} $+1$ &
\multicolumn{1}{|c}{$0$} & \multicolumn{1}{|c|}{$-1$} \\
\cline{2-3}
\end{tabular}
\end{center}
\caption{A potential function of the $3$-player minority game}
\label{fig:potential}
\end{figure}
The Nash equilibria of the three-player game follow easily from
the results in Section~\ref{sec:equilibrium}. Throughout this
section, Nash equilibria are denoted by $(p,q,r) \in [0,1]^3$,
where $p,q,r$ are the probabilities with which player $1$, $2$,
and $3$, respectively, choose $-1$. Then, the Nash equilibria of
the game are $(1/2,1/2,1/2)$ and $(1, 0, \lambda)$ for some
$\lambda \in [0,1]$, and permutations of these.

Given parameter $\beta \geq 0$, the conditions for a logit quantal
response equilibrium (QRE) become:
\begin{eqnarray}
p & = & \frac{1}{1 + \exp - \beta (1 - q - r)}, \label{eq:Cp} \\
q & = & \frac{1}{1 + \exp - \beta (1 - p - r)}, \label{eq:Cq} \\
r & = & \frac{1}{1 + \exp - \beta (1 - p - q)}. \label{eq:Cr}
\end{eqnarray}
Given $\beta \geq 0$, we denote a logit QRE in which player 1,2
and 3 play $-1$ with probability $p,q,r$ by $(p,q,r,\beta)$. We
now characterize the set of Nash equilibria that are the limit of
a sequence of quantal response equilibria when $\beta \to \infty$.
\begin{proposition} \label{prop:limiting_QRE}
Let $(p(\beta_n), q(\beta_n), r(\beta_n), \beta_n)_{n \in
\mathbb{N}}$ be a sequence of logit quantal response equilibria:
$\beta_n \rightarrow \infty$ and for each $n \in \mathbb{N}$, the
quadruple $(p(\beta_n), q(\beta_n), r(\beta_n), \beta_n)$ solves
equations~\eqref{eq:Cp}-\eqref{eq:Cr}. A Nash equilibrium
$(p,q,r)$ is the limit of such a sequence if and only if one of
the following conditions hold:
\begin{description}
\item[(a)] $(p,q,r)$ is a pure Nash equilibrium,

\item[(b)] $(p,q,r)$ is a Nash equilibrium with exactly one mixer
who mixes uniformly,

\item[(c)] $(p,q,r) = (1/2,1/2,1/2)$.
\end{description}
\end{proposition}
\noindent The proof is in Appendix~\ref{app:limiting_QRE}.
Proposition~\ref{prop:limiting_QRE} thus characterizes the set of
stationary points of the perturbed best response dynamics
\eqref{eq:PV} for the three-player minority game.

\section{Best-reply learning with limited memory} \label{sec:L2BP}

In this section, we consider discrete time learning models in
which players choose best replies to beliefs that are supported by
observed play in the recent past. We study two such models, the
learning model proposed by \citet{Hurkens1995} and the model of
\citet{KetsVoorneveld2005}. First, in the learning model of
\citeauthor{Hurkens1995}, players may choose any action that is a
best reply to some belief over other players' actions that is
consistent with their recent past play. The limiting behavior of
this learning process is easy to characterize.
\citeauthor{Hurkens1995} shows that the Markov processes defined
by his learning process eventually settle down in so-called
minimal curb sets \citep{BasuWeibull1991}. Minimal curb sets are
product sets of pure strategies containing all best responses
against beliefs restricted to the recommendations to the remaining
players. Unfortunately, this does not provide a sharp prediction
in the minority game. As shown by \citet{TercieuxVoorneveld2005},
the unique minimal curb set in the minority game consists of the
entire strategy space. That is, over time, all players will keep
on choosing both actions.

Secondly, we study the model of \citet{KetsVoorneveld2005}. As in
the model of \citet{Hurkens1995}, it is assumed that players
best-respond to beliefs over others' play supported by recent past
play. In addition, players display a so-called recency bias: when
there are multiple best replies to a given belief, a player
chooses the best reply that he most recently played.\footnote{The
behavioral economics literature provides several motivations for
the common observation that agents appear somewhat unwilling to
deviate from their recent choices. This can be attributed to e.g.
the formation of habits \citep[cf.][]{Young1998} or the use of
rules of thumb \citep[cf.][]{EllisonFudenberg1993}.}
\citeauthor{KetsVoorneveld2005} show that play converges to one of
the minimal prep sets of the game under this learning process.
Minimal prep sets \citep{Voorneveld2004} are a set-valued solution
concept for strategic games that combines a standard rationality
condition, stating that the set of recommended strategies to each
player must contain at least one best reply to whatever belief he
may have that is consistent with the recommendations to the other
players, with players' aim at simplicity, which encourages them to
maintain a set of strategies that is as small as possible. Think
of the set of recommendations to a player in a minimal prep set as
a well-packed suitcase for a holiday: you want to be prepared for
different kinds of weather, but bringing all five of your
umbrellas and all seven bathing suits may be overdoing it.
\citet{TercieuxVoorneveld2005} show that the minimal prep sets of
the minority game and the pure Nash equilibria of the game
coincide. Hence, under the learning model of
\citeauthor{KetsVoorneveld2005}, play in the minority game
converges to one of the pure Nash equilibria of the game.

In both the model of \citet{Hurkens1995} and
\citet{KetsVoorneveld2005}, players need to recall a sufficiently
long period of play in order for play to converge. We now turn to
the question what this lower bound on players' memory is. More
specifically, suppose players remember actions that were chosen
during the past $T \in \mathbb{N}$ periods. A memory length of $T
= 1$ is clearly insufficient for a best-reply learning process
with limited memory to converge. If players chose an action
profile yesterday that is not a pure Nash equilibrium, then some
action, say $-1$, was chosen by more than $k+1$ players. Hence,
everyone chooses the unique best reply $+1$ today, and
consequently the unique best reply $-1$ to this tomorrow, and the
unique best reply $+1$ to this the day after tomorrow, with action
profiles forever cycling between these two extremes. 
However, we show that a memory length $T = 2$ is sufficient for
the learning process of \citet{KetsVoorneveld2005} to convergence
to pure Nash equilibria.

When the memory length $T $ is equal to $2$, the process is a
Markov chain with state space $H = \{(a^1,a^2) \mid a^1, a^2 \in
A^{2k+1}\}$, where a \emph{history} $h = (a^1, a^2) \in H$
indicates that the $2k+1$ players remember that they chose action
profile $a^1$ one period ago and $a^2$ two periods ago. Having
defined the set $H$ of states, we proceed to the \emph{transition
probability functions} $P: H \times H \rightarrow [0,1]$, where
$P(h,h') \in [0,1]$ is the probability of moving from state $h \in
H$ to state $h' \in H$ in one period and $\sum_{h' \in H} P(h,h')
= 1$ for all $h \in H$. We do not need to specify exact
probabilities: for the convergence result, only sign restrictions
are needed.

Moving from $h = (a^1,a^2)$ to $h' = (b^1,b^2)$ in one period
means that $h'$ is obtained from $h$ after one more round of play,
i.e., by appending a new profile of most recent actions. Formally:
\begin{itemize}
\item[\textbf{[P1]}] $h' = (b^1,b^2)$ is a successor of $h =
(a^1,a^2)$, i.e., $b^2 = a^1$.
\end{itemize}
Moreover, by moving from $h = (a^1,a^2)$ to $h' = (b^1,b^2)$, the
processes in \citet{KetsVoorneveld2005} require that each player
$i \in N$ chooses a best reply to a belief $\alpha_{-i} \in
\times_{j \in N \setminus \{i\}} \Delta (\{a^1_j, a^2_j\})$ with
support in the product set of actions chosen in the previous $T =
2$ periods, whenever possible sticking to the most recent best
reply. In games with just two actions, the latter simply means
that you continue playing as you did in the previous round, unless
that action is no longer a best reply to your current belief.
Formally:
\begin{itemize}
\item[\textbf{[P2]}] For each $i \in N$, $b^1_i$ is a best reply
to some belief $\alpha_{-i} \in \times_{j \in N \setminus \{i\}}
\Delta (\{a^1_j, a^2_j\})$. Moreover, $b^1_i = a^1_i$ if and only
if $a^1_i$ is a best reply to $\alpha_{-i}$.
\end{itemize}

\begin{proposition}
Consider a Markov chain on $H$ with transition probability
function $P$, where, for all states $h,h' \in H$, it holds that
$P(h,h') > 0$ if and only if [P1] and [P2] are true. This Markov
process eventually settles down in a pure Nash equilibrium.
\end{proposition}
\begin{proofw}
Let $h_0 = (a^1, a^2) \in H$ and distinguish two cases:

\noindent \textsc{Case 1: $a^1$ is a pure Nash equilibrium.} By
[P2], the players will react with positive probability to the
belief that everybody plays as in $a^1$. Each player's most recent
best reply is to continue playing as in $a^1$, so the process
moves with positive probability to the history $h_1 = (a^1, a^1)$.
From here on, the only feasible belief based on the past two
periods is that the players play $a^1$ and the most recent best
reply implies that they will continue to play $a^1$: the process
stays in state $h_1$ and play has converged to a pure Nash
equilibrium.

\noindent \textsc{Case 2: $a^1$ is not a pure Nash equilibrium.}
By Proposition~\ref{prop:pure_NE}, some alternative, w.l.o.g.
$-1$, was chosen by a set $S \subseteq N$ of players with $|S| >
k+1$. Each player's unique best response to $a^1$ is therefore to
choose $+1$. By [P2], the process moves with positive probability
to state $h_1 = ((+1, \ldots, +1), a^1)$. Let $a^* \in A^{2k+1}$
be a pure Nash equilibrium where $k+1$ members of $S$ choose $+1$
and the others choose $-1$. Again using [P2], the process moves
with positive probability from $h_1$ to $h_2 = (a^*, (+1, \ldots,
+1))$:
\begin{itemize}
\item For each of the selected $k+1$ members of $S$, $+1$ is the
unique best reply to the belief drawn from the past two periods
that at least $k+1$ other players from $S$ will choose $-1$.

\item For each of the remaining $k$ players, $-1$ is the unique
best response to the belief that all other players will continue
to play last period's profile $(+1, \ldots, +1)$.
\end{itemize}
Notice that history $h_2$ belongs to case 1.

Conclude that, regardless of the initial state $h_0$, the Markov
process moves with positive probability to an absorbing state
where the players continue to play one of the game's pure Nash
equilibria. As the Markov process is finite and the initial state
was chosen arbitrarily, this will eventually happen with
probability one \citep{KemenySnell1976}: play eventually settles
down in a pure Nash equilibrium.
\end{proofw}

\medskip
\noindent Some remarks are in order. First, notice that, due to
the symmetry of the minority game, a minor revision of the proof
indicates that convergence to pure Nash equilibria can be
established also if the only thing players remember from the past
two periods is what they chose themselves and \emph{how many
others\/} did so. This comes at the expense of a more complex
notation and a larger deviation from that of
\citet{KetsVoorneveld2005}.

Secondly, the result that the lower bound on players' memory
length is two indicates that the requirement on memory length in
\citet{KetsVoorneveld2005} for general games can be decreased
significantly in specific cases. Although the convergence result
in \citet{KetsVoorneveld2005} for the entire class of finite
strategic games also applies here, we include an explicit proof:
the structure of a minority game allows us to give a considerably
shorter proof of the convergence result for this specific game,
and allows us to derive a much sharper bound on the memory length.

\section{Concluding remarks} \label{sec:concl}

Though congestion games are apparently simple, game-theorists'
understanding of play in such games is far from complete, for two
reasons. Firstly, well-known learning models do not always provide
equivocal predictions for such games. In this paper, we have
characterized the Nash equilibria and the limiting behavior of
several well-known learning models in a simple congestion game. We
show that these learning models provide different predictions.
Secondly, experimental results are not always in line with
theoretical predictions. In experiments on market entry games,
aggregate play is largely consistent with equilibrium play, with
the number of entrants close to capacity, but individual play
generally does not resemble Nash play \citep[see
e.g.][]{Ochs1999}. Hence, an interesting direction for future
research would be to test behavior in minority games
experimentally. This provides the opportunity to compare the
performance of different learning models in the minority
game.\footnote{\citet{BottazziDevetag2007} and
\citet{ChmuraPitz2006} present experiments on the minority game.
However, their results cannot be directly used to compare the
performance of different learning models, as they do not test
explicitly whether play converges to particular strategy profiles
or to particular product sets of actions. Both papers merely study
the effect of information on players' aggregate payoffs.}
Moreover, it may help to better understand behavior in other
congestion games such as the market entry games, as the symmetry
of the game makes it harder for players to play repeated-game
strategies. In experiments on the (asymmetric) market entry games,
players sometimes seem to follow such strategies, with some
players choosing to enter the market in every round in the initial
periods, regardless of payoffs, to obtain a reputation for always
entering \citep[see][for a discussion]{DuffyHopkins2005}. Such
strategies are useless in the minority game, so that it may be
hoped that the minority game offers a cleaner test of the theory.



\appendix
\makeatletter
\def\@seccntformat#1{\csname Pref@#1\endcsname \csname the#1\endcsname\quad}
\def\Pref@section{Appendix~}
\makeatother

\section{Stochastic dominance for binomial distributions} \label{app:stoch_dom_binom}

Let $X$ have a binomial distribution with $n \in \mathbb{N}$ draws
and success probability $p \in [0,1]$; briefly, a $B(n,p)$
distribution: $X = X_1 + \cdots + X_n$, where $X_1, \ldots, X_n$
are i.i.d $B(1,p)$. Distributions with a higher success rate $p$
\emph{stochastically dominate\/} those with a lower one
\citep[cf.][Exc. 9.9]{Ross1996}. Formally, in terms of cumulative
distributions, if $p, q \in [0,1]$ and $p < q$, then
\[
\mbox{For all } m \in \{0, \ldots, n\}: \quad \sum_{k = 0}^m {n
\choose k} p^k (1-p)^{n-k} \geq \sum_{k = 0}^m {n \choose k} q^k
(1-q)^{n-k},
\]
with strict inequality if $m < n$. This follows by substitution if
$m = 0$ or $m = n$. So let $m \in \{1, \ldots, n-1\}$. It suffices
to show that the function
\[
[0,1] \ni p \mapsto \sum_{k=0}^m {n \choose k} p^k (1-p)^{n-k}
\]
has a negative derivative on $(0,1)$. The derivative, after
rewriting, becomes
\begin{eqnarray*}
\lefteqn{\sum_{k=0}^m {n \choose k} \left[k p^{k-1} (1-p)^{n-k} -
(n-k) p^k (1-p)^{n-k-1} \right]} \\
&& = \sum_{k=0}^m {n \choose k} p^{k-1}(1-p)^{n-k-1} \left[ k - np \right] \\
&& = \sum_{k=0}^m {n \choose k} k p^{k-1} (1-p)^{n-k-1} - n
\sum_{k=0}^m {n \choose k} p^k (1-p)^{n-k-1} \\
&& = {n \over 1-p} \sum_{k=0}^{m-1} {n-1 \choose k} p^k
(1-p)^{n-1-k} - {n \over 1-p} \sum_{k=0}^m {n \choose k} p^k
(1-p)^{n-k} \\
&& = {n \over 1-p} \left[ \mathbb{P}\left(\sum_{k=1}^{n-1} X_k
\leq m-1 \right) - \mathbb{P}\left(\sum_{k=1}^{n} X_k \leq m
\right) \right].
\end{eqnarray*}
Consider the term in square brackets. The first probability is
strictly smaller than the second, as the first event (at most
$m-1$ successes in the first $n-1$ draws) implies the second one
(at most $m$ successes during all $n$ draws), whereas the latter
also includes the positive-probability event that
$\sum_{k=1}^{n-1} X_k = m$. Hence, the derivative is negative, as
we had to show.

Write a function $g: \{0, 1, \ldots, n\} \rightarrow \mathbb{R}$
as the sum of indicator functions: 
\begin{eqnarray*}
g & = & g(n) \mathbb{I}_{\{0, \ldots, n\}} + (g(n-1) - g(n))
\mathbb{I}_{\{0, \ldots, n-1\}} + \cdots + (g(0) - g(1))
\mathbb{I}_{\{0\}} \\
& = & g(n) \mathbb{I}_{\{0, \ldots, n\}} + \sum_{k=0}^{n-1} (g(k)
- g(k+1)) \mathbb{I}_{\{0, \ldots, k\}}.
\end{eqnarray*}
Then
\[
\mathbb{E}[g(X)] = g(n) + \sum_{k=0}^{n-1} (g(k) - g(k + 1))
\mathbb{P}(X \leq k).
\]
If $g$ is nonconstant, nonincreasing, then $g(k) - g(k+1) \geq 0$
for all $k = 0, \ldots, n-1$, with at least one strict inequality.
As shown above, the cumulative probabilities are strictly
decreasing in the success probability $p$. So $\mathbb{E}[g(X)]$
becomes a strictly decreasing function of $p$: the higher the
probability of success, the larger the probability that $g(X)$
achieves a low value. Of course, for nondecreasing functions the
converse holds.

\section{Proof of Proposition~\ref{prop:limiting_QRE}}
\label{app:limiting_QRE}

The only Nash equilibria not covered by (a), (b), and (c) are
those with one player (w.l.o.g. player 1) choosing $-1$, one
player (w.l.o.g. player 2) choosing $+1$, and the third player
(w.l.o.g. player 3) mixing with probability $\lambda \in (0,1)
\setminus \{\tfrac{1}{2}\}$.

Suppose, to the contrary, that such an equilibrium is the limit of
a sequence of logit QRE $(p(\beta_n), q(\beta_n), r(\beta_n),
\beta_n)_{n \in \mathbb{N}}$ where $\beta_n \rightarrow \infty$
and $(p(\beta_n), q(\beta_n), r(\beta_n), \beta_n)$ solves
equations~\eqref{eq:Cp} to \eqref{eq:Cr} for a logit QRE. In the
selected equilibrium, both the $(-1)$-player and the $(+1)$-player
choose their unique best response. By Lemma 3 in \citet[][p.
251]{Turocy2005}, $\beta_n(1-p(\beta_n)) \rightarrow 0$ and
$\beta_n q(\beta_n) \rightarrow 0$. Substituting this in the logit
QRE condition~\eqref{eq:Cr} for the third player gives that
$$
r(\beta_n) = {1 \over 1 + \exp -\beta_n(1 - p(\beta_n) -
q(\beta_n))} \rightarrow \frac{1}{2},
$$
contradicting the assumption that $\lim_{n \rightarrow \infty}
r(\beta_n) = \lambda \neq 1/2$.

\bigskip

\noindent It remains to show that the classes of equilibria in the
proposition are indeed limits of a sequence of logit QREs.

\medskip
\noindent \textbf{(a):} By symmetry, it suffices to show that the
pure Nash equilibrium $(p,q,r) = (1,1,0)$ is the limit of a
sequence of logit QREs.

\noindent \textbf{Step 1:} For each $\beta > 4$ there is a logit
QRE $(p,q,r,\beta)$ with $p=q \in (1/2,1)$, and $r < 1/2$.

\noindent \textbf{Proof of Step 1:} Based on
conditions~\eqref{eq:Cp} - \eqref{eq:Cr} for a logit QRE and the
substitution $p = q$, define for all $\beta > 0$ and $p \in
[1/2,1]$:
\begin{gather*}
r(p,\beta) := \frac{1}{1 + \exp[-\beta(1-2p)]},\\
f(p,\beta) := \frac{1}{1 + \exp[-\beta(1-p-r(p,\beta))]}.
\end{gather*}
Let $\beta > 4$. We show that there is a solution $p^* \in
(1/2,1]$ to the equation $p = f(p,\beta)$. Substitution in
\eqref{eq:Cp} - \eqref{eq:Cr} yields that $(p,q,r,\beta) =
(p^*,p^*,r(p^*,\beta),\beta)$ is a logit QRE with the desired
properties. Notice that
\begin{eqnarray*}
\frac{\partial f(p,\beta)}{\partial p} &=& \frac{-\beta
\exp\left(-\beta (1-p - r(p, \beta))\right)}{\left(1+
\exp\left(-\beta (1-p - r(p, \beta))\right)\right)^2} \left(1 + \frac{\partial r(p,\beta)}{\partial p}\right),\\
&=& \frac{-\beta \exp\left(-\beta (1-p - r(p,
\beta))\right)}{\left(1+ \exp\left(-\beta (1-p -
r(p,\beta_0))\right)\right)^2} \left(1 + \frac{-2\beta \exp\left(
-\beta(1-2p)\right)}{\left(1 + \exp\left( -\beta(1-2p)\right)
\right)^2}\right) .
\end{eqnarray*}
Since $f(1/2,\beta) = 1/2$ and
$$\frac{\partial
f(1/2,\beta)}{\partial p} = -
\frac{\beta}{4}\left(\frac{2-\beta}{2}\right) >1$$ for $\beta >
4$, it follows that $f(p,\beta) > p$ for $p$ slightly larger than
$1/2$. Moreover, $f(1,\beta) <1$. Hence, by the Intermediate Value
Theorem applied to $f(\cdot, \beta)$, $f(p^*,\beta) = p^*$ for
some $p^* \in (1/2,1)$.

\noindent \textbf{Step 2:} Let $\beta_0 > 4$ and let $p_0 \in
(1/2,1)$ solve $f(p_0,\beta_0) = p_0$. This is possible by Step 1.
The function $f(p_0, \cdot)$ is strictly increasing on $[\beta_0,
\infty)$.

\noindent \textbf{Proof of Step 2:} By definition of $f$, it
suffices to show that the derivative of
\[
\beta \mapsto \beta (1 - p_0 - r(p_0, \beta)), \qquad \beta \in
[\beta_0, \infty)
\]
is positive. This derivative equals
\begin{equation}\label{eq:derivative}
- \beta \frac{\partial r(p_0,\beta)}{\partial \beta} + 1 - p_0 -
r(p_0, \beta).
\end{equation}
Using $p_0 > 1/2$ and the definition of $r$, it follows that
$\partial r(p_0,\beta)/\partial \beta < 0$, i.e., the function
$r(p_0, \cdot)$ is strictly decreasing on $[\beta_0, \infty)$.
Moreover, as $p_0 = f(p_0,\beta_0) > 1/2$, it follows from the
definition of $f$ that $1 - p_0 - r(p_0, \beta_0) > 0$. As $r(p_0,
\cdot)$ is decreasing, this implies that $1 - p_0 - r(p_0, \beta)
> 0$ for each $\beta \in [\beta_0, \infty)$. Therefore, the
expression in \eqref{eq:derivative} is positive.

\noindent \textbf{Step 3:} The pure Nash equilibrium $(p,q,r) =
(1,1,0)$ is the limit of a sequence of QREs.

\noindent \textbf{Proof of Step 3:} Let $\beta_0 > 4$ and consider
a QRE $(p_0,q_0,r_0,\beta_0)$ as in Step 1. Set $\beta_1 = \beta_0
+ 1$. By Step 2, $p_0 = f(p_0,\beta_0) < f(p_0,\beta_1)$.
Moreover, $f(1,\beta_1) < 1$. By the Intermediate Value Theorem
applied to the function $f(\cdot, \beta_1)$, there is a $p_1 \in
(p_0,1)$ with $p_1 = f(p_1,\beta_1)$. Conclude that there is a QRE
$(p_1,q_1,r_1,\beta_1)$ with
\[
\begin{array}{cclcccc}
p_1 & = & q_1 & = & f(p_1, \beta_1) & > & p_0, \\
r_1 & = & r(p_1,\beta_1) &&&&\\
\beta_1 & = & \beta_0 + 1 &&&&
\end{array}
\]
Repeating this construction allows us to define a sequence
$(p_n,q_n,r_n,\beta_n)_{n \in \mathbb{N}}$ of solutions to
\eqref{eq:Cp} - \eqref{eq:Cr} satisfying the conditions of Step 1
and with $\beta_n \rightarrow \infty$ and $(p_n)_{n \in
\mathbb{N}}$ strictly increasing.

As $(p_n,q_n,r_n)_{n \in \mathbb{N}}$ is a sequence in the compact
strategy space, we may assume w.l.o.g. that the sequence
converges. Its limit $(p,q,r)$ must be a Nash equilibrium
\citep{McKelveyPalfrey1995}. As $(p_n)_{n \in \mathbb{N}}$ is a
strictly increasing sequence in $(1/2,1)$ and $(r_n)_{n \in
\mathbb{N}}$ is a sequence in $(0,1/2)$, it must be $p=q > 1/2$
and $r \leq 1/2$. The only Nash equilibrium of the game with these
properties is $(p,q,r) = (1,1,0)$.

\medskip
\noindent \textbf{(b):} By symmetry, it suffices to show that the
Nash equilibrium $(p,q,r) = (1,0,1/2)$ is the limit of a sequence
of logit QREs. The steps are similar to those in (a). Therefore,
the proof is kept short.

\medskip
\noindent \textbf{Step 1:} For each $\beta > 4$ there is a logit
QRE $(p,q,r,\beta)$ with $p \in (1/2,1), q = 1 - p, r=1/2$.

\noindent \textbf{Proof of Step 1:} Let $\beta > 4$. Based on the
substitution $q = 1 - p$ and $r = 1/2$ in condition \eqref{eq:Cp}
for a logit QRE, define
$$
g(p, \beta) := \frac{1}{1 + \exp \left[ \beta (1/2 - p)\right]}.
$$
We show that there is a solution $p^* \in (1/2,1)$ to the equation
$p = g(p,\beta)$. Substitution in \eqref{eq:Cp} - \eqref{eq:Cr}
yields that $(p,q,r,\beta) = (p^*,1-p^*,1/2,\beta),\beta)$ is a
logit QRE with the desired properties. Notice that
$$
\frac{\partial g(p, \beta)}{\partial p} = \frac{\beta \exp\left[
\beta (1/2 - p)\right]}{(1 + \exp \beta (1/2 - p))^2}.
$$
Since $g(1/2, \beta) = 1/2$ and $\partial g(1/2, \beta)/\partial p
= \beta/4>1$, it follows that $g(p, \beta) > p$ for $p$ slightly
larger than $1/2$. Moreover, $g(1,\beta) < 1$, so the Intermediate
Value Theorem implies that $g(p^*, \beta) = p^*$ for some $p^* \in
(1/2, 1)$.

\noindent \textbf{Step 2:} For each $p_0 \in (1/2,1)$, the
function $g(p_0, \cdot)$ is strictly increasing on $(0, \infty)$.

\noindent \textbf{Proof of Step 2:} Immediate from the definition
of $g$.

\noindent \textbf{Step 3:} The Nash equilibrium $(p,q,r) =
(1,0,1/2)$ is the limit of a sequence of logit QREs.

\noindent \textbf{Proof of Step 3:} Reasoning as in the proof of
step 3 in part (a) allows us to construct a sequence
$(p_n,q_n,r_n,\beta_n)_{n \in \mathbb{N}}$ of solutions to
\eqref{eq:Cp} - \eqref{eq:Cr} satisfying the conditions of Step 1
and with $\beta_n \rightarrow \infty$ and $(p_n)_{n \in
\mathbb{N}}$ strictly increasing. As $(p_n,q_n,r_n)_{n \in
\mathbb{N}}$ is a sequence in the compact strategy space, we may
assume w.l.o.g. that the sequence converges. Its limit $(p,q,r)$
must be a Nash equilibrium \citep{McKelveyPalfrey1995}. As
$(p_n)_{n \in \mathbb{N}}$ is a strictly increasing sequence in
$(1/2,1)$, $q_n = 1 - p_n$ and $r_n = 1/2$ for all $n \in
\mathbb{N}$, it must be $p > 1/2, q = 1-p, r= 1/2$. The only Nash
equilibrium of the game with these properties is $(p,q,r) =
(1,0,1/2)$.

\medskip
\noindent \textbf{(c):} It follows by substitution that $(p,q,r,
\beta) = (1/2, 1/2, 1/2, \beta)$ is a logit QRE for all $\beta
\geq 0$. Consequently, the Nash equilibrium $(p,q,r) =
(1/2,1/2,1/2)$ is the limit of a sequence of logit QREs with
$\beta \to \infty$.

\bibliographystyle{chicago}
\bibliography{bibl_GMG}

\begin{thebibliography}{}

\bibitem[\protect\citeauthoryear{Arthur}{Arthur}{1994}]{Arthur1994}
Arthur, W.~B. (1994).
\newblock Inductive reasoning and bounded rationality.
\newblock {\em American Economic Review\/}~{\em 84}, 406--411.

\bibitem[\protect\citeauthoryear{Basu and Weibull}{Basu and
  Weibull}{1991}]{BasuWeibull1991}
Basu, K. and J.~Weibull (1991).
\newblock Strategy subsets closed under rational behavior.
\newblock {\em Economics Letters\/}~{\em 36}, 141--146.

\bibitem[\protect\citeauthoryear{Bena{\"i}m}{Bena{\"i}m}{1999}]{Benaim1999}
Bena{\"i}m, M. (1999).
\newblock Dynamics of stochastic algorithms.
\newblock In J.~Az{\'e}ma, M.~Emery, M.~Ledoux, and M.~Yor (Eds.), {\em
  S{\'e}minaire de Probabilit{\'e}s XXXIII}, Volume 1709 of {\em Lecture Notes
  in Mathematics}, pp.\  1 -- 68. Berlin: Springer-Verlag.

\bibitem[\protect\citeauthoryear{Blonski}{Blonski}{1999}]{Blonski1999}
Blonski, M. (1999).
\newblock Anonymous games with binary actions.
\newblock {\em Games and Economic Behavior\/}~{\em 28}, 171--180.

\bibitem[\protect\citeauthoryear{Bottazzi and Devetag}{Bottazzi and
  Devetag}{2007}]{BottazziDevetag2007}
Bottazzi, G. and G.~Devetag (2007).
\newblock Competition and coordination in experimental minority games.
\newblock {\em Journal of Evolutionary Economics\/}~{\em 17}, 241 -- 275.

\bibitem[\protect\citeauthoryear{Challet, Marsili, and Zhang}{Challet
  et~al.}{2004}]{ChalletMarsiliZhang2004}
Challet, D., M.~Marsili, and Y.-C. Zhang (2004).
\newblock {\em Minority Games: Interacting Agents in Financial Markets}.
\newblock Oxford: Oxford University Press.

\bibitem[\protect\citeauthoryear{Chmura and Pitz}{Chmura and
  Pitz}{2006}]{ChmuraPitz2006}
Chmura, T. and T.~Pitz (2006).
\newblock Successful strategies in repeated minority games.
\newblock {\em Physica A\/}~{\em 363}, 477 -- 480.

\bibitem[\protect\citeauthoryear{Coolen}{Coolen}{2005}]{Coolen2005}
Coolen, A. (2005).
\newblock {\em The Mathematical Theory of Minority Games: Statistical Mechanics
  of Interacting Agents}.
\newblock Oxford: Oxford University Press.

\bibitem[\protect\citeauthoryear{DeAngelis and Gross}{DeAngelis and
  Gross}{1992}]{DeAngelisGross1992}
DeAngelis, D. and L.~Gross (1992).
\newblock {\em Individual-Based Models and Approaches in Ecology: Populations,
  Communities, and Ecosystems}.
\newblock New York: Chapman and Hall.

\bibitem[\protect\citeauthoryear{Duffy and Hopkins}{Duffy and
  Hopkins}{2005}]{DuffyHopkins2005}
Duffy, J. and E.~Hopkins (2005).
\newblock Learning, information, and sorting in market-entry games: theory and
  evidence.
\newblock {\em Games and Economic Behavior\/}~{\em 51}, 31--62.

\bibitem[\protect\citeauthoryear{Ellison and Fudenberg}{Ellison and
  Fudenberg}{1993}]{EllisonFudenberg1993}
Ellison, G. and D.~Fudenberg (1993).
\newblock Rules of thumb for social learning.
\newblock {\em Journal of Political Economy\/}~{\em 101}, 612 -- 643.

\bibitem[\protect\citeauthoryear{Franke}{Franke}{2003}]{Franke2003}
Franke, R. (2003).
\newblock Reinforcement learning in the {El} {Farol} model.
\newblock {\em Journal of Economic Behavior and Organization\/}~{\em 51},
  367--388.

\bibitem[\protect\citeauthoryear{Harsanyi and Selten}{Harsanyi and
  Selten}{1988}]{HarsanyiSelten1988}
Harsanyi, J.~C. and R.~Selten (1988).
\newblock {\em A General Theory of Equilibrium Selection in Games}.
\newblock Cambridge, MA: MIT Press.

\bibitem[\protect\citeauthoryear{Helbing, Sch{\"o}nhof, Stark, and
  Holust}{Helbing et~al.}{2005}]{HelbingSchoenhofStark2005}
Helbing, D., M.~Sch{\"o}nhof, H.-U. Stark, and J.~A. Holust (2005).
\newblock How individuals learn to take turns: {E}mergence of alternating
  cooperation in a congestion game and the {P}risoners {D}ilemma.
\newblock {\em Advances in Complex Systems\/}~{\em 8}, 87 -- 116.

\bibitem[\protect\citeauthoryear{Hofbauer and Hopkins}{Hofbauer and
  Hopkins}{2005}]{HofbauerHopkins2005}
Hofbauer, J. and E.~Hopkins (2005).
\newblock Learning in perturbed asymmetric games.
\newblock {\em Games and Economic Behavior\/}~{\em 52}, 133--152.

\bibitem[\protect\citeauthoryear{Hofbauer and Sandholm}{Hofbauer and
  Sandholm}{2002}]{HofbauerSandholm2002}
Hofbauer, J. and W.~H. Sandholm (2002).
\newblock On the global convergence of stochastic fictitious play.
\newblock {\em Econometrica\/}~{\em 70}, 2265--2294.

\bibitem[\protect\citeauthoryear{Hopkins}{Hopkins}{2002}]{Hopkins2002}
Hopkins, E. (2002).
\newblock Two competing models of how people learn in games.
\newblock {\em Econometrica\/}~{\em 70}, 2141--2166.

\bibitem[\protect\citeauthoryear{Huberman and Lukose}{Huberman and
  Lukose}{1997}]{HubermanLukose1997}
Huberman, B. and R.~Lukose (1997).
\newblock Social dilemmas and internet congestion.
\newblock {\em Science\/}~{\em 277}, 535--537.

\bibitem[\protect\citeauthoryear{Hurkens}{Hurkens}{1995}]{Hurkens1995}
Hurkens, S. (1995).
\newblock Learning by forgetful players.
\newblock {\em Games and Economic Behavior\/}~{\em 11}, 304--329.

\bibitem[\protect\citeauthoryear{Kemeny and Snell}{Kemeny and
  Snell}{1976}]{KemenySnell1976}
Kemeny, J.~G. and J.~L. Snell (1976).
\newblock {\em Finite Markov Chains}, Volume~40 of {\em Undergraduate Texts in
  Mathematics}.
\newblock Berlin: Springer-Verlag.

\bibitem[\protect\citeauthoryear{Kets and Voorneveld}{Kets and
  Voorneveld}{2005}]{KetsVoorneveld2005}
Kets, W. and M.~Voorneveld (2005).
\newblock Learning to be prepared.
\newblock {\em SSE/EFI Working Paper\/}~{\em 590}.

\bibitem[\protect\citeauthoryear{Kojima and Takahashi}{Kojima and
  Takahashi}{2004}]{KojimaTakahashi2004}
Kojima, F. and S.~Takahashi (2004).
\newblock Anti-coordination games and dynamic stability.
\newblock {\em Working Paper, Harvard University\/}.
\newblock Forthcoming in International Game Theory Review.

\bibitem[\protect\citeauthoryear{McKelvey and Palfrey}{McKelvey and
  Palfrey}{1995}]{McKelveyPalfrey1995}
McKelvey, R.~D. and T.~R. Palfrey (1995).
\newblock Quantal response equilibria for normal form games.
\newblock {\em Games and Economic Behavior\/}~{\em 10}, 6--38.

\bibitem[\protect\citeauthoryear{Meyer, Van~Huyck, Battalio, and Saving}{Meyer
  et~al.}{1992}]{Meyer_et_al_1992}
Meyer, D.~J., J.~B. Van~Huyck, R.~C. Battalio, and T.~R. Saving (1992).
\newblock History's role in coordinating decentralized allocation decisions.
\newblock {\em Journal of Political Economy\/}~{\em 100}, 292--316.

\bibitem[\protect\citeauthoryear{Monderer and Shapley}{Monderer and
  Shapley}{1996}]{MondererShapley1996}
Monderer, D. and L.~Shapley (1996).
\newblock Potential games.
\newblock {\em Games and Economic Behavior\/}~{\em 14}, 124--143.

\bibitem[\protect\citeauthoryear{Nagel, Rasmussen, and Barrett}{Nagel
  et~al.}{1997}]{NagelRasmussenBarrett1997}
Nagel, K., S.~Rasmussen, and C.~Barrett (1997).
\newblock Network traffic as a self-organized critical phenomenon.
\newblock In F.~Schweitzer (Ed.), {\em Self-organization of Complex Structures:
  From Individual to Collective Dynamics}, pp.\  579. London: Gordon and
  Breach.

\bibitem[\protect\citeauthoryear{Ochs}{Ochs}{1999}]{Ochs1999}
Ochs, J. (1999).
\newblock Coordination in market entry games.
\newblock In D.~Budescu, I.~Erev, and R.~Zwick (Eds.), {\em Games and Human
  Behavior: Essays in Honor of Amnon Rapoport}, pp.\  143--172. Mahwah, NJ:
  Erlbaum.

\bibitem[\protect\citeauthoryear{Renault, Scarlatti, and Scarsini}{Renault
  et~al.}{2005}]{RenaultScarlattiScarsini2005}
Renault, J., S.~Scarlatti, and M.~Scarsini (2005).
\newblock A folk theorem for minority games.
\newblock {\em Games and Economic Behavior\/}~{\em 53}, 208 -- 230.

\bibitem[\protect\citeauthoryear{Rosenthal}{Rosenthal}{1973}]{Rosenthal1973}
Rosenthal, R. (1973).
\newblock A class of games possessing pure-strategy {Nash} equilibria.
\newblock {\em International Journal of Game Theory\/}~{\em 2}, 65--67.

\bibitem[\protect\citeauthoryear{Ross}{Ross}{1996}]{Ross1996}
Ross, S. (1996).
\newblock {\em Stochastic processes\/} (Second ed.).
\newblock New York, NY: Wiley \& Sons.

\bibitem[\protect\citeauthoryear{Sandholm}{Sandholm}{2001}]{Sandholm2001}
Sandholm, W.~H. (2001).
\newblock Potential games with continuous player sets.
\newblock {\em Journal of Economic Theory\/}~{\em 97}, 81--108.

\bibitem[\protect\citeauthoryear{Sandholm}{Sandholm}{2007}]{Sandholm2007}
Sandholm, W.~H. (2007).
\newblock {\em Population Games and Evolutionary Dynamics}.
\newblock Cambridge, MA: MIT Press.
\newblock Forthcoming.

\bibitem[\protect\citeauthoryear{Selten and G{\"u}th}{Selten and
  G{\"u}th}{1982}]{SeltenGuth1982}
Selten, R. and W.~G{\"u}th (1982).
\newblock Equilibrium point selection in a class of market entry games.
\newblock In M.~Diestler, E.~F{\"u}rst, and G.~Schw{\"o}diauer (Eds.), {\em
  Games, Economic Dynamics, and Time Series Analysis}. Wien-W{\"u}rzburg:
  Physica-Verlag.

\bibitem[\protect\citeauthoryear{Tercieux and Voorneveld}{Tercieux and
  Voorneveld}{2005}]{TercieuxVoorneveld2005}
Tercieux, O. and M.~Voorneveld (2005).
\newblock The cutting power of preparation.
\newblock {\em SSE/EFI Working Paper\/}~{\em 583}.

\bibitem[\protect\citeauthoryear{Turocy}{Turocy}{2005}]{Turocy2005}
Turocy, T. (2005).
\newblock A dynamic homotopy interpretation of the logistic quantal response
  equilibrium correspondence.
\newblock {\em Games and Economic Behavior\/}~{\em 51}, 243 -- 263.

\bibitem[\protect\citeauthoryear{Voorneveld}{Voorneveld}{2004}]{Voorneveld2004}
Voorneveld, M. (2004).
\newblock Preparation.
\newblock {\em Games and Economic Behavior\/}~{\em 48}, 403 -- 414.

\bibitem[\protect\citeauthoryear{Weibull}{Weibull}{1995}]{Weibull1995}
Weibull, J.~W. (1995).
\newblock {\em Evolutionary Game Theory}.
\newblock Cambridge, MA: MIT Press.

\bibitem[\protect\citeauthoryear{Young}{Young}{1998}]{Young1998}
Young, H.~P. (1998).
\newblock {\em Individual Strategy and Social Structure}.
\newblock Princeton, NJ: Princeton University Press.

\end{thebibliography}
\end{document}